\pdfoutput=1
\documentclass[reprint,12pt,onecolumn,notitlepage,nofootinbib,floatfix]{revtex4-2}
\usepackage{amssymb}
\usepackage{amsmath}
\usepackage{amsfonts} 
\usepackage{bm}
\usepackage{color} 
\usepackage{graphicx}
\usepackage{xcolor}

\allowdisplaybreaks

\usepackage{epsfig} 
\usepackage{epstopdf}
\usepackage{psfrag}   
\usepackage{subfigure}  
\usepackage{booktabs}  
\usepackage{braket}
\usepackage{mathtools}
\usepackage{textcomp}
\usepackage{ifthen}
\usepackage{tensor}
\usepackage{protosem} 
\usepackage{wasysym}

\usepackage[toc]{appendix}
\usepackage{color,soul} 
\usepackage{datetime}
\usepackage{hyperref}
\hypersetup{pdfstartview=FitH,pdfhighlight=/O,colorlinks=true}
\definecolor{darkblue}{rgb}{0.2, 0, 0.8}
\definecolor{darkgreen}{rgb}{0.2, 0.71, 0}
\definecolor{awesome}{rgb}{1.0, 0.13, 0.32}
\definecolor{cadmiumred}{rgb}{0.89, 0.0, 0.13}
\definecolor{dukeblue}{rgb}{0.0, 0.0, 0.61}

\newcommand{\bea}{\begin{eqnarray}}
\newcommand{\eea}{\end{eqnarray}}
\newcommand{\ba}{\begin{eqnarray}}
\newcommand{\ea}{\end{eqnarray}}

\newcommand{\beq}{\begin{equation}}
\newcommand{\eeq}{\end{equation} }
\newcommand{\beqa}{\begin{eqnarray}}
\newcommand{\eeqa}{\end{eqnarray}}
\newcommand{\beqar}{\begin{eqnarray*}}
\newcommand{\eeqar}{\end{eqnarray*}}
\newcommand{\e}{\epsilon}

\newcommand{\D}{\mathcal{D}}

\let\a=\alpha \let\b=\beta   \let\e=\varepsilon
   
     \let\r=v

  \let\D=\Delta

\let\pa=\partial
\newcommand{\Eqref}[1]{Eq.~\eqref{#1}}

\newcommand{\figref}[1]{Fig.~(\ref{#1})}

\renewcommand{\(}{\left(}
\renewcommand{\)}{\right)}

\newcommand{\bpa}{\bar{\partial}}

\renewcommand{\href}[2]{#2}

\usepackage{setspace}
\usepackage{etoolbox}
\makeatletter
\patchcmd{\@footnotetext}
 {\setspace@singlespace}{0.8}
 {}{}
\makeatother

\begin{document}

\title{Do holographic CFT states have unique semiclassical bulk duals?}

\author{Stefano Antonini}
\email{Corresponding author, santonini@berkeley.edu}

\author{Pratik Rath}
\email{pratik\_rath@berkeley.edu}
\affiliation{Center for Theoretical Physics and Department of Physics,
University of California, Berkeley, CA 94720, USA}

\date{\today}

\begin{abstract}  \vskip 0.2in 

We discuss a situation where a holographic CFT state has multiple semiclassical bulk duals. 
In our example, a given holographic state has two simple, semiclassical descriptions, one with a closed universe, constructed using the gravitational path integral, and one without a closed universe, constructed using the extrapolate dictionary.
This highlights an ambiguity in the AdS/CFT dictionary.
We propose various options for resolving this tension although none are perfectly satisfactory.
We also discuss what this implies for the black hole interior and the gravitational path integral.

\vskip 0.2in
\centering
\noindent {\it \footnotesize Essay written for the Gravity Research Foundation 2025 Awards for Essays on Gravitation}

\end{abstract}   

\maketitle 

\newpage

\section{Introduction}

The AdS/CFT duality was a remarkable development in our understanding of quantum gravity \cite{Maldacena:1997re}.
While it led to strong evidence in favour of black hole evaporation being unitary, the existence of the black hole interior remained controversial \cite{Mathur:2009hf,Almheiri:2012rt,Almheiri:2013hfa}.
Recent progress has led to the derivation of the Page curve, demonstrating unitarity of black hole evaporation, by using the gravitational Euclidean path integral  \cite{Penington:2019npb,Almheiri:2019psf,Penington:2019kki,Almheiri:2019qdq}.
This demonstrates the power of semiclassical methods in answering difficult questions that were previously thought to require a complete description of quantum gravity.
Nevertheless, new questions are raised by these calculations.

An important ingredient of these recent calculations is that the inner product is corrected by wormhole contributions. Thus, naively orthogonal semiclassical configurations may be non-orthogonal, at least as CFT states.
These wormhole contributions then lead to the factorization problem, i.e. they are in tension with the computation of fine-grained quantities in AdS/CFT unless either a) the bulk theory is dual to an ensemble of boundary theories or b) additional contributions to the path integral exist that cancel their effect.\footnote{In option b), wormholes correctly capture the statistics of coarse-grained CFT quantities \cite{Belin:2020hea,Chandra:2022bqq,Sasieta:2022ksu,deBoer:2023vsm}.}
The first option is well understood in 2D gravity and by now many examples exist starting from the work of Ref.~\cite{Saad:2019lba}.
However, ensemble averaging of interacting, holographic CFTs in higher dimensions is difficult to understand owing to the sparseness expected of conformal fixed points at finite central charge.
The second option is realized by half-wormholes in toy models \cite{Saad:2021rcu,Mukhametzhanov:2021nea,Mukhametzhanov:2021hdi,Gesteau:2024gzf}, whereas we currently have no such understanding of the new bulk ingredients in more general theories.

These facts were reformulated in Ref.~\cite{Marolf:2020xie} in the language of the Hilbert space of closed universes.
Assuming an inner product defined using the gravitational path integral that includes wormholes, Ref.~\cite{Marolf:2020xie} showed that there is a one-to-one correspondence between the gauge-inequivalent bulk states of closed universes and the different theories in the boundary ensemble that are being averaged over.\footnote{This argument requires an independent, UV-complete definition of bulk quantum gravity via a bulk path integral as well as a formulation using an AdS/CFT-type dictionary relating bulk and boundary partition functions. However, it need not be restricted to a local integral over the metric and other fields, it can include new ingredients like strings, branes etc. The precise set of assumptions is enlisted in Ref.~\cite{Colafranceschi:2023urj}.}
In the standard AdS/CFT case of a single boundary theory, this means that all the different closed universe states are secretly gauge equivalent.
This is often stated as the fact that the Hilbert space of closed universes is one-dimensional.

While this choice of inner product is quite attractive from the perspective of the boundary CFT, it is unclear whether it is relevant for the experience of a bulk observer.
Ref.~\cite{Akers:2022qdl} made an attempt to reconcile these wormhole corrections with the experience of a bulk observer in the black hole interior.
They argued that while the map from the semiclassical bulk Hilbert space to the boundary Hilbert space is many-to-one, and thus non-isometric, it is in fact invertible when restricted to low complexity states.
Thus, they argued that the bulk observer experiences usual semiclassical physics at least as long as no high complexity operations are performed.
In particular, the inner products with and without wormholes differ only in exponentially small corrections and for the experience of the bulk observer, one may as well use either.
Nevertheless, it leaves open the question of which inner product is relevant for the bulk observer more generally, for instance in typical black hole microstates which are high-complexity, or at very late times in an evaporating black hole.
For the most part, we will assume that the inner product includes wormholes as would be a natural option for the AdS/CFT duality, although we will explore alternate options at the end.

In order to fully understand the experience of a bulk observer, apart from the relevant inner product, we also need to figure out the dictionary relating boundary CFT states to bulk states.
For simple low energy states, the extrapolate dictionary provides a unique bulk description.
However, as mentioned above, in the presence of black holes we generally have naively orthogonal bulk semiclassical descriptions that secretly have non-zero overlap with each other as CFT states. This can lead to a non-trivial kernel for the bulk-to-boundary map \cite{Akers:2022qdl,chao_et_al:LIPIcs.ITCS.2017.48,Cao:2023gkw}.
Given a particular state in the CFT, it then appears that one could have many different bulk representations of the state although some are complicated superpositions of semiclassical descriptions.

In this essay, we will discuss an example where the different bulk representations are in fact semiclassical.
This is achieved by focusing our attention on closed universes.
In particular, we present an example of a state with two simple semiclassical descriptions: one description contains a closed universe and the other does not.
We will then discuss possible interpretations of this result, what this means for the black hole interior, and the implications for the gravitational path integral.

\enlargethispage{0.7\baselineskip}

\section{Closed Universes and Multiple Bulk Descriptions}\label{closed}

In order to build a setup with multiple semiclassical bulk descriptions, we will consider a CFT state whose bulk dual includes a closed universe.
How does one construct closed universes in AdS/CFT? 
The first construction came from the Maldacena-Maoz type wormhole \cite{Maldacena:2004rf} where one finds geometries connecting two different asymptotic AdS boundaries. The bulk state prepared on the reflection symmetric slice separating the two boundaries is that one of a closed, crunching universe.
Such wormholes can be stabilized by adding matter, see e.g. Refs.~\cite{VanRaamsdonk:2020tlr,VanRaamsdonk:2021qgv,Antonini:2022blk,Antonini:2022ptt,Chandra:2022bqq,Usatyuk:2024mzs} among others.
However, if the Euclidean saddle only involves a Maldacena-Maoz wormhole, there is no boundary state being prepared and it is difficult to understand how to interpret the bulk physics unless one has an ensemble of boundary theories \cite{Marolf:2020xie}.

Instead, one can try to construct CFT states whose dual description involves non-gravitational CFT degrees of freedom entangled with closed universes. This can be done using bra-ket wormhole constructions \cite{Chen:2020tes,VanRaamsdonk:2020tlr,VanRaamsdonk:2021qgv,Antonini:2022blk,Antonini:2022ptt}, in which the closed universe is an entanglement island \cite{Almheiri:2019hni,Hartman:2020khs,Bousso:2022gth} for, and therefore in the entanglement wedge of, the non-gravitational CFT.
Similarly, whenever one has the interior of a black hole as an island, a canonical purification of the state of the radiation results in a closed universe entangled with the radiation \cite{Chandrasekaran:2020qtn,Akers:2022max}. 
In either of these cases, there is a large amount of radiation entangled with the closed universe and the state of the radiation is quite complex. 
Therefore, the arguments of Ref.~\cite{Akers:2022qdl} go through and the description of the closed universe is semiclassical up to exponentially (and parametrically) small corrections in the entanglement.
We will instead be interested in situations where the amount of entanglement is small in order to sharpen the tension.

\begin{figure}
    \centering
    \includegraphics[scale=0.45]{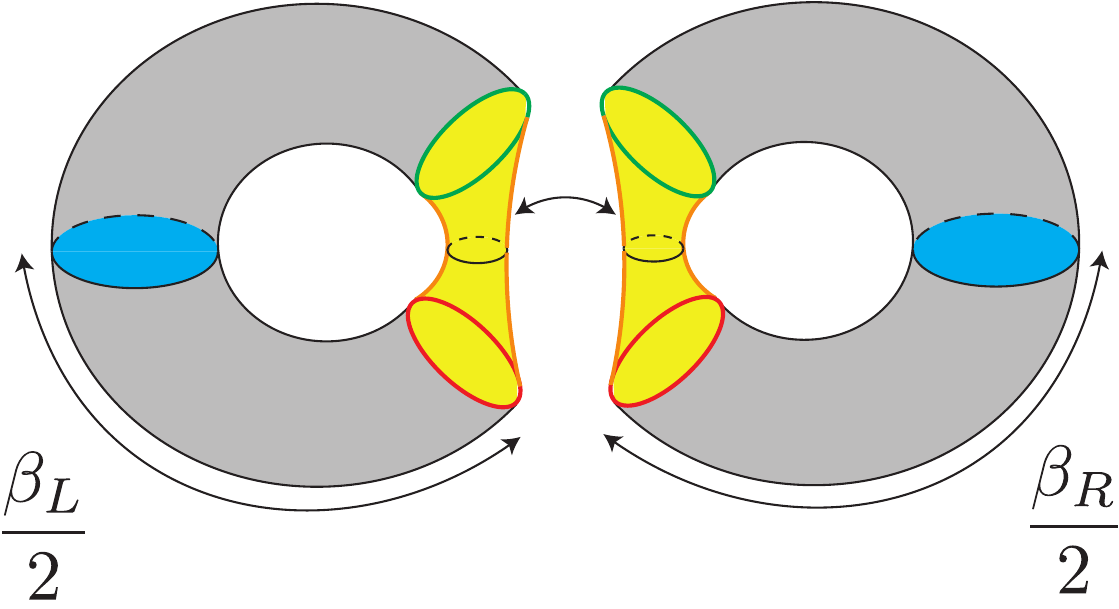}
    \caption{The Euclidean saddle computing $\braket{\Psi_i|\Psi_i}$ is built up from two thermal AdS saddles by gluing the boundaries of the yellow regions (orange lines) to each other. The two green (red) lines---i.e. operator insertions in the bra (ket)---are identified. The jump in extrinsic curvature is supported by the matter shock produced by the operator insertions in the bra and ket. The time symmetric slice contains two AdS spaces (blue) and a closed universe (yellow) giving us Description 1 \cite{Antonini:2023hdh}.}
    \label{D1}
\end{figure}

We will analyze a setup discussed in Ref.~\cite{Antonini:2023hdh}.
For concreteness, we will work in AdS$_3$/CFT$_2$ although the same arguments apply in higher dimensions.
Consider a partially entangled thermal state (PETS) \cite{Goel:2018ubv,Balasubramanian:2022gmo}, namely an entangled state of two CFTs similar to the thermofield double state, where one adds a heavy, spherically symmetric operator insertion $\mathcal{O}^{(i)}$ in the Euclidean past:
\begin{equation}
    \ket{\psi_i\(\b_L,\b_R\)} = \frac{1}{\sqrt{Z}}\sum_{n,m} e^{-\frac{1}{2}(\b_L E_n +\b_R E_m)}\mathcal{O}^{(i)}_{nm}\ket{E_n^*}_L\ket{E_m}_R,
\end{equation}
where heavy means the conformal dimension $\Delta_{\mathcal{O}^{(i)}}$ is above the black hole threshold and $i$ labels different possible choices of matter insertion.\footnote{The different choices $i$ need not be related by any symmetry, but could for example be associated with different scaling dimensions.}
For large $\b_L,\b_R$, we expect the state to effectively be below the Hawking-Page transition.
By demanding an approximate global symmetry under which the bulk dual of $\mathcal{O}^{(i)}$ is charged, we can ensure that $\mathcal{O}^{(i)}$ has a vanishing one-point function and thus the insertions in the bra and ket must be paired. 
This leads to the dominant Euclidean saddle being of the form shown in \figref{D1}.\footnote{It would be interesting to improve this saddle by finding smooth solutions supported by smeared matter fields although we see no reason to suspect that the qualitative features would change.}
The time-symmetric slice suggests that the bulk dual of the state $\ket{\psi_i\(\b_L,\b_R\)}$ consists of two copies of AdS with radiation that is entangled with a closed universe supported by the matter insertion \cite{Antonini:2023hdh}.
We call this Description 1. Note that in Description 1, we have $\braket{\Psi_i|\Psi_j}=0$ for $i\neq j$, whereas $|\braket{\Psi_i|\Psi_j}|^2\sim e^{-S}$ due to a wormhole contribution, where $S$ is the $O(1)$ entanglement entropy of the closed universe with the two copies of AdS \cite{Antonini:2023hdh}.

For convenience, we will truncate the state on both CFTs to energies within a microcanonical window of size $\D_0\gg1$ but still parametrically $O(1)$. 
We can choose $\D_0$ large enough to ensure most of the wavefunction is captured by it and the truncated tail has a probability weight no greater than $\e$.\footnote{We consider a microcanonical version of the setup in Ref.~\cite{Antonini:2023hdh} to avoid any subtlety discussed there involving thermal tails, e.g. the possibility that the peculiarities of specific closed universe states are encoded in the features of the tails of the specific PETS.}
Since we have chosen the window to include the average ADM energy of the saddle, the bulk geometry remains unchanged at leading order after adding in such energy projectors $P_{\D_0}$.\footnote{This can be made precise by using an inverse Laplace transform a la Ref.~\cite{Marolf:2018ldl} to project onto the microcanonical window.} The effect of $P_{\D_0}$ is simply to truncate the thermal tails of the bulk state of radiation.

Now the puzzle arises because this is a state of $O(1)$ energy (in AdS units) and $O(1)$ entropy.
By the extrapolate dictionary, the bulk description of such states is well understood.
To be concrete, consider a bulk theory whose low energy matter content is just a single scalar field $\phi$ whose boundary dual is a single trace operator $\Phi$.
By the state-operator correspondence in the boundary CFT, the local operators of $O(1)$ conformal dimension provide a complete basis for the CFT states of $O(1)$ energy on a circle.
In this example, the only operators of $O(1)$ conformal dimension would be of the form $\pa^{l_1} \bpa^{k_1} \Phi \pa^{l_2} \bpa^{k_2} \Phi \dots \pa^{l_n} \bpa^{k_n} \Phi$, namely primaries and descendents of $\Phi$ and its multi-trace counterparts.
For simplicity, we will label these operators $\Phi_j$.

\begin{figure}
    \centering
    \includegraphics[scale=0.45]{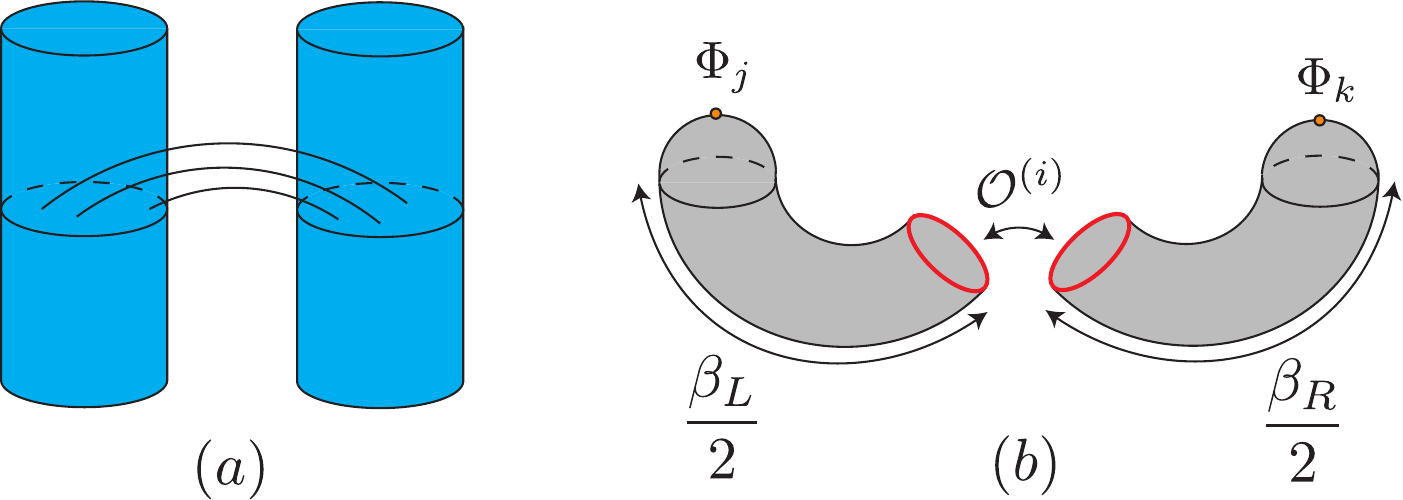}
    \caption{a) Description 2 consists of two AdS spaces (blue) that are entangled with each other (represented by the lines connecting them). b) The coefficients of the wavefunction $c_{jk}^{(i)}$ can be computed using the CFT path integral on a Riemann surface with operator insertions $\Phi_{j,k}$ and $\mathcal{O}^{(i)}$.}
    \label{D2}
\end{figure}

With this complete low-energy basis in hand, we can expand the state of interest in this basis as
\begin{equation}\label{eq:sum}      \ket{\Psi_i}=P_{\D_0,L}P_{\D_0,R}\ket{\psi_i\(\b_L,\b_R\)}= \sum_{j,k} c^{(i)}_{jk} \Phi_j^{(L)} \Phi_k^{(R)}|0\rangle_L|0\rangle_R,
\end{equation}
where the sum is restricted to operators with conformal dimensions within the microcanonical window.
The amplitudes $c^{(i)}_{jk}$ can be then computed as
\begin{equation}
        c^{(i)}_{jk}=\langle 0|_L\langle 0|_R \Phi_j^{(L)} \Phi_k^{(R)} e^{-\beta_L H_L} \mathcal{O}^{(i)} e^{-\beta_R H_R} |max\rangle,
        \label{eq:coeff}
\end{equation}
where $\ket{max}$ is the maximally entangled state on the two CFTs.
Thus, we have expressed these coefficients in terms of a light-light-heavy (LLH) three point function on a particular Riemann surface, see \figref{D2}.
Such LLH three point functions have been studied in Ref.~\cite{Pappadopulo:2012jk} for instance and are known to be appropriately pseudorandom.
In effect, we end up with $\ket{\Psi_i}$ being a low energy state with a pseudorandom wavefunction.
Nevertheless, we are imagining that we have complete control of the holographic CFT we are simulating and can completely determine these OPE coefficients to prepare precisely the state $\ket{\Psi_i}$.

Now we can also use the fact that \Eqref{eq:sum} has a standard bulk dual using the extrapolate dictionary relating the CFT operator $\Phi$ to the bulk scalar field $\phi$ \cite{Banks:1998dd}.
This means that we can entirely re-express \Eqref{eq:sum} as a bulk state involving bulk operators acting on the product of two AdS vacuums. 
In this description, there is no closed universe and the bulk entanglement is purely between radiation in these two asymptotically AdS universes, see \figref{D2}.
We call this Description 2. In Description 2, the bulk inner product $\braket{\Psi_i|\Psi_j}$ between different states is by construction given by their (non-zero) inner product in the CFT.
Given the precise coefficients $c^{(i)}_{jk}$, this bulk description is semiclassical in the sense that it is completely captured by quantum field theory on curved spacetime. Note, however, that the values of the exact coefficients $c_{jk}^{(i)}$ should be regarded as a UV input. In fact, it is difficult to imagine a semiclassical Euclidean saddle being able to compute the fine-grained pseudorandom coefficients, and it instead only captures coarse grained quantities (which behave like averages over the pseudorandom ensemble) via wormholes \cite{Belin:2020hea,Chandra:2022bqq,Sasieta:2022ksu,deBoer:2023vsm}. This explains the behavior of the inner product in Description 1.

We now have two semiclassical bulk descriptions of the same CFT state and this makes the bulk-boundary dictionary ambiguous.\footnote{This ambiguity is of the same type pointed out in Ref.~\cite{Marolf:2012xe} except we have an explicit semiclassical Euclidean saddle constructing the closed universe. A similar puzzle in the evaporating black hole context was also pointed out and discussed in \cite{Bousso:2019ykv,Bousso:2020kmy}.}
Moreover, the state lives in a subspace of dimension $O(1)$, hence its complexity is also at most $O(1)$.
Thus, no complexity criteria can resolve the issue.
So how can we interpret this result?
Below we discuss some possibilities.

\paragraph{Ensemble Averaging}: 
A natural fix would be if the gravitational path integral actually computes an average over an ensemble of dual theories.
In this case, note that Description 1 was constructed in the Hartle Hawking (HH) state since we assumed the presence of no other boundaries in the gravitational path integral.
On the other hand, Description 2 used the precise OPE coefficients of a single CFT, which on the bulk side corresponds to an $\a$-state.
Such $\a$-states are expected to be coherent states of closed universes.
This can be seen precisely in simple models such as the topological model of Ref.~\cite{Marolf:2020xie}.
One can then imagine that in an $\a$-state, the closed universes lying around can absorb the closed universe found in Description 1 to produce a bulk with just two copies of AdS.
The two descriptions then correspond to distinct states in the bulk Hilbert space and can very well coexist.

Of course, the down side of this option is that we only have an understanding of ensemble averaging in two dimensional theories like JT gravity.
In higher dimensional theories, it is difficult to find enough CFTs to ensemble average over.
The lack of an ensemble has also been argued on the basis of the swampland cobordism conjecture \cite{McNamara:2020uza}.
On the other hand, some progress towards a definition of a dual ensemble in higher dimensions have been made \cite{Colafranceschi:2023urj,Belin:2023efa,Jafferis:2024jkb}.

\paragraph{No Semiclassical Closed Universe}:
If we insist on there being no ensemble averaging, then Description 2 would be preferred as an effective description of the CFT state, because the extrapolate dictionary is a basic tenet of AdS/CFT.
In any case, the Euclidean saddle with a closed universe is a poor effective theory for computing CFT observables.
As we have discussed, for semiclassically orthogonal states, we have $\braket{\Psi_i|\Psi_j}=\delta_{ij}$, whereas $|\braket{\Psi_i|\Psi_j}|^2\sim e^{-S}$ for $i\neq j$ due to a wormhole contribution, where $S$ is $O(1)$ \cite{Antonini:2023hdh}.
This is an avatar of the factorization problem.
If we further assume that the inner product relevant for the bulk observer in the closed universe is the same as the CFT inner product, this would mean the closed universe is non-semiclassical.
The Euclidean saddle only captures coarse-grained averages, but observables in the closed universe have large fluctuations.
These large corrections to the physics of the closed universe must arise from hitherto unknown effects such as half-wormholes \cite{Saad:2021rcu,Mukhametzhanov:2021nea,Mukhametzhanov:2021hdi,Gesteau:2024gzf}.
When such effects are included, it is plausible that they yield the precise state of Description 2 from a bulk perspective, although the details are far from understood.

In summary, this option means that, assuming a single dual CFT exists and its inner product is the one relevant for a bulk observer, the naively semiclassical looking closed universe is actually not semiclassical.
One could say it lies in the swampland of gravitational states.
Assuming AdS/CFT is a complete description of quantum gravity, the downside of this option is obviously that semiclassical closed universes would not exist in a theory of quantum gravity.
This would be a stark prediction since we could apriori have been living in a closed universe.
A potential loophole though is that even if $O(10)$ qubits are entangled with the asymptotic universes, the errors from wormhole effects are much smaller than the precision of experiments that we have performed to date.
This in fact would be the natural extrapolation of the results of Ref.~\cite{Akers:2022qdl}.

\paragraph{Beyond AdS/CFT}:
If we assume that a bulk observer in a closed universe should experience ordinary semiclassical physics\footnote{This is a well-motivated assumption: there is no conclusive evidence that we do not live in a closed universe \cite{DiValentino:2019qzk,Handley:2019tkm,Planck:2018vyg}, and our local experience should be independent of the global properties of the universe.} and that it should be possible to describe a closed universe in a UV-complete theory of quantum gravity, we would have to conclude that AdS/CFT is not a complete description of bulk quantum gravity.
There would then be a more complete bulk description able to capture an observer's experience, which only agrees with AdS/CFT when restricted to global questions asked from a CFT point of view.
This, of course, would also be quite drastic, although interesting such scenarios have been proposed recently \cite{Marolf:2024jze,Abdalla:2025gzn}.
Ref.~\cite{Marolf:2024jze} proposed that the bulk theory may potentially violate reflection-positivity in the presence of a large number of asymptotic boundaries without affecting the semiclassical physics of closed universes.
Of course, the CFT, being reflection positive, is then unable to capture any such physics. Ref.~\cite{Abdalla:2025gzn}, on the other hand, proposed a new way to describe quantum gravity in the presence of an observer using the gravitational path integral, and concluded that the Hilbert space relevant to describe the observer's experience is much larger than the Hilbert space of the CFT living on the asymptotic AdS boundary.

A toy example that helps understand option d) is to think of reconstructing the entanglement wedge \cite{Dong:2016eik} of a subregion $R$ of the CFT.
Consider a situation where $R$ has two candidate entanglement wedges $W_{1,2}$ (with $W_1\subset W_2$) with generalized entropies that differ only by an $O(1)$ amount such as the example of Ref.~\cite{Akers:2019wxj}.
In this case, just like in our closed universe example, the entanglement wedge is not sharply defined in the sense that there are $O(1)$ errors in computing overlaps of the state in the bulk region $W_2\backslash W_1$ by computing overlaps of the CFT density matrix on subregion $R$.
This can be seen by computing the fidelity of the Petz map \cite{Penington:2019kki}.
This may lead one to conclude that semiclassical physics breaks down for the bulk region $W_2\backslash W_1$ by arguments analogous to the ones we made above.
However, in this case global reconstruction is completely well-defined and there is an obvious inner product that is preferred for describing the semiclassical physics of this bulk region: the inner product of the entire CFT.
Our example could be analogous to this, with the difference that $O(1)$ errors arise even when performing global reconstruction. It may be the case that the closed universe is semiclassical with its usual inner product but its non-perturbative completion is something beyond even the entire CFT.

The following flowchart summarizes the three possibilities outlined above.
\begin{figure}[h]
    \centering
    \includegraphics[width=0.8\textwidth]{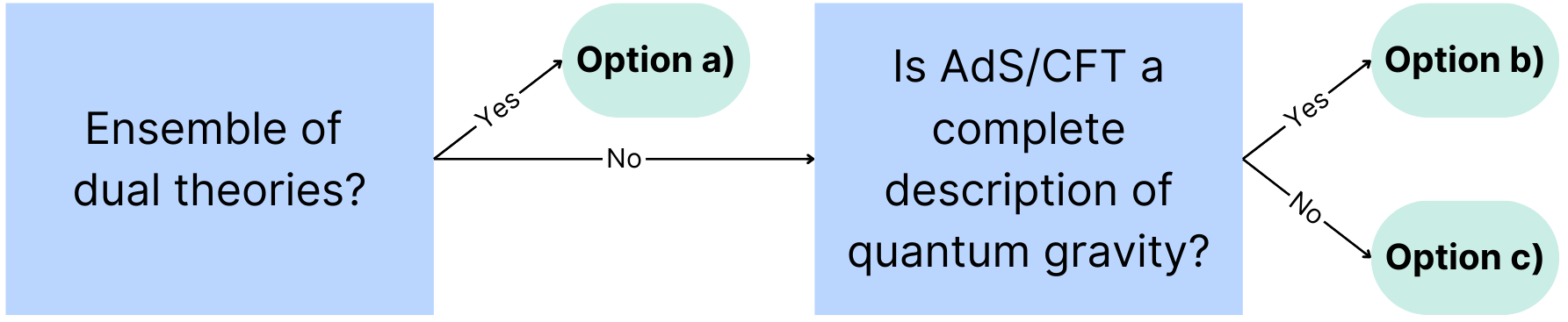}
\end{figure}

\vspace{-.8cm}

\section{Implications for the Black Hole Interior}\label{BH}

Having discussed our particular example, we could ask what this means for the black hole interior.
At sufficiently late times when the black hole is much (but not parametrically) larger than the Planck scale as seen from outside, the setup is very similar to our closed universe setup. Description 1 corresponds to a black hole with a smooth, large interior highly entangled to Hawking radiation, whereas Description 2 corresponds to the exterior radiation being nearly pure.
One can then apply each of our proposed options to this setting, which in fact has been discussed in the literature in some form or the other.

Ref.~\cite{Marolf:2020xie} and, with a slightly different approach, Refs.~\cite{Bousso:2019ykv,Bousso:2020kmy} proposed that the existence of an ensemble makes the calculations of the Page curve consistent.
Of course, one may then ask what happens if we calculate the Page curve in an $\a$ state.
In this case, if we extend the analogy from the closed universe, the black hole interior will be non-semiclassical, i.e. it stops existing.
This also continues to be the case if we use option b). A bulk picture compatible with option b) is also provided by the final state proposal of Ref.~\cite{Horowitz:2003he}, in which semiclassical physics in the black hole interior breaks down (see \cite{Bousso:2013uka} for a discussion of various issues with this proposal). 
Option c) on the other hand says that the black hole interior potentially exists, but AdS/CFT cannot tell us anything about it.

\vspace{-.3cm}

\section{Implications for the gravitational path integral}

Finally, we would like to comment briefly about the implications of the three options outlined above for the role of the path integral in quantum gravity.

If option a) is correct and the dual description is given by an ensemble of dual boundary theories, the path integral in the saddle point approximation computes quantities averaged over such an ensemble. This is compatible with the behavior of the overlaps in Description 1 described above. In the language of Ref.~\cite{Marolf:2020xie}, Description 1 captures physics in the Hartle-Hawking state, whereas Description 2 represents a specific $\alpha$-state in the baby universe Hilbert space.

\begin{figure}
    \centering
    \includegraphics[width=\linewidth]{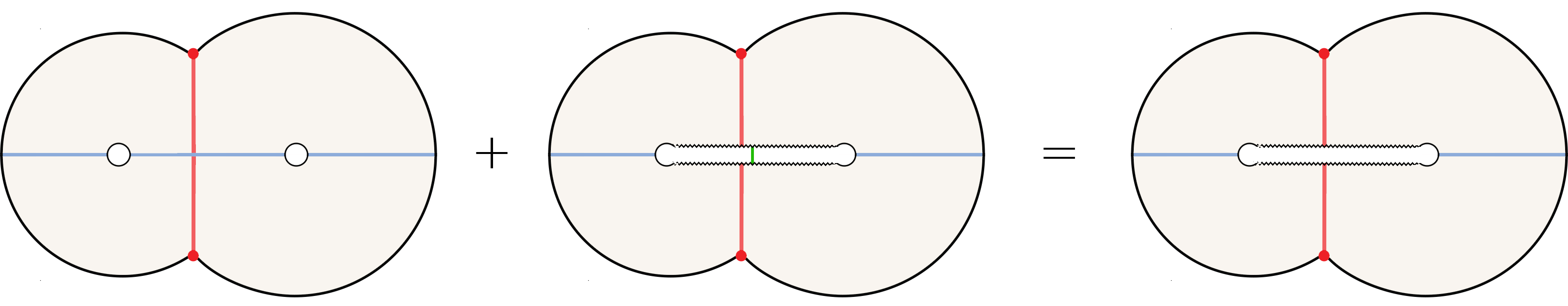}
    \caption{Pictorial representation of how the non-perturbative gravitational path integral could yield Description 2. When additional, non-perturbative ingredients are taken into account, ``half-wormhole''-like saddles (center figure) can contribute to the path integral \cite{Saad:2021rcu,Mukhametzhanov:2021nea,Mukhametzhanov:2021hdi,Gesteau:2024gzf}, with the closed universe replaced by some non-semiclassical correlation. The sum of such saddles and the semiclassical saddle associated with Description 1 (left figure) could be equivalent to a Euclidean geometry without a closed universe (right figure), which would lead to Description 2.}
    \label{fig:halfwormholes}
\end{figure}

If option b) is correct, then the path integral, at least in the saddle point approximation, is only able to capture coarse-grained quantities, which behave like averages over an ensemble due to the pseudorandom statistics of the coefficients $c_{jk}^{(i)}$ defined in equation \eqref{eq:coeff} \cite{Belin:2020hea,Chandra:2022bqq,Sasieta:2022ksu,deBoer:2023vsm}. An interesting question is whether the full, non-perturbative path integral is able to compute fine-grained quantities, which should factorize, and yield Description 2. This was shown to be possible in toy models, where half-wormholes  rescue factorization \cite{Saad:2021rcu,Mukhametzhanov:2021nea,Mukhametzhanov:2021hdi,Gesteau:2024gzf}. It seems plausible that additional ingredients (such as strings, branes, and other UV objects) need to be included in the path integral in order to obtain this result, as it was the case in Refs. \cite{Saad:2021rcu,Mukhametzhanov:2021nea,Mukhametzhanov:2021hdi,Gesteau:2024gzf}, see Figure \ref{fig:halfwormholes}.

Finally, option c) could imply that the gravitational path integral, at least the way it is usually defined, consistently computes bulk observables only when they correspond to well-defined observables in the boundary CFT. In particular, such observables should act linearly on the CFT Hilbert space. When they do not---as it is the case, e.g., for the experience of an observer in a closed universe or infalling into an old black hole, when observables become state-dependent and therefore non-linear on the CFT Hilbert space---a different definition of the path integral might be needed. For instance, this new definition could involve taking into account the presence of an observer as proposed in \cite{Abdalla:2025gzn}, or violating reflection positivity as proposed in \cite{Marolf:2024jze}, or it might not exist at all. This would not be shocking, given the well-known difficulties in defining the bulk quantum gravity path integral in higher dimensions independently of a dual boundary path integral (see \cite{Gibbons:1978ac,Marolf:2022ybi} and references therein).

In summary, we have used a simple example to restate the confusions that continue to be associated with the black hole information puzzle as well as with the physics of closed universes.
The key point we would like to emphasize to the readers is that we need a deeper knowledge of the AdS/CFT dictionary in order to resolve the tension we posed, understand physics in the black hole interior and in closed universes, and elucidate the precise role of the gravitational path integral in the definition of quantum gravity.

\vskip 0.2in
\emph{Acknowledgments:}
We are thankful to Alex Belin, Raphael Bousso, Xi Dong, Daniel Harlow, Thomas Hartman, Adam Levine, Javier Magan, Don Marolf, Henry Maxfield, Shiraz Minwalla, Geoff Penington, Mukund Rangamani, Phil Saad, Martin Sasieta, Arvin Shahbazi-Moghaddam, Brian Swingle, and Herman Verlinde for useful conversations related to the topics of the essay. This work was supported in part by the Berkeley Center for Theoretical Physics; and by the Department of Energy, Office of Science, Office of High Energy Physics under QuantISED Award DE-SC0019380.  S.A. is supported by the U.S.
Department of Energy through DE-FOA-0002563. 


\renewcommand{\leftmark}{\MakeUppercase{Bibliography}}
\phantomsection
\bibliographystyle{JHEP}
\bibliography{references}
\label{biblio}

\end{document}